\documentclass[twocolumn,showpacs,preprintnumbers,amsmath,amssymb,aps]{revtex4-2}

\usepackage{epsfig}
\usepackage{dcolumn}
\usepackage{bm}
\usepackage[dvipsnames]{xcolor}
\usepackage[normalem]{ulem}
\usepackage{natbib,hyperref}

\usepackage{multirow}
\usepackage{adjustbox}

\begin{document}

\preprint{\today} 

\title{Measurement of the hyperfine coupling constants and absolute energies of the $8p \ ^2P_{1/2}$ and $8p \ ^2P_{3/2}$ levels in atomic cesium}

\author{Jonah A. Quirk$^{1,2}$, Liam Sherman$^{1}$, Amy Damitz$^{1,2}$, Carol E. Tanner$^3$ and D. S. Elliott$^{1,2,4}$}

\affiliation{%
   $^1$Department of Physics and Astronomy, Purdue University, West Lafayette, Indiana 47907, USA\\
   $^2$Purdue Quantum Science and Engineering Institute, Purdue University, West Lafayette, Indiana 47907, USA\\
   $^3$Department of Physics and Astronomy, University of Notre Dame, Notre Dame, Indiana 46556, USA\\
    $^4$School of Electrical and Computer Engineering, Purdue University, West Lafayette, Indiana 47907, USA
   }

\date{\today}

\begin{abstract}
We report measurements of the hyperfine coupling constant for the $8p \ ^2P_{1/2}$ level of atomic cesium, $^{133}$Cs, with a relative uncertainty of $\sim$0.019\%. Our result is $A = 42.933 \: (8)$ MHz, in good agreement with recent theoretical results. We also examine the hyperfine structure of the $8p \ ^2P_{3/2}$ state, and derive new values for the state energies of the $8p \ ^2P_{1/2}$ and $8p \ ^2P_{3/2}$ states of cesium.
\end{abstract}
\maketitle 

\section{Introduction}\label{sec:introduction}
Atomic parity violation (APV) measurements provide a window through which the weak-force interaction between nucleons and electrons at low-collision energies can be viewed.  The weak-force interaction perturbs the atomic system, rendering optical transitions that would otherwise be strictly forbidden slightly allowed.  The amplitude for these interactions is weak, typically $\sim10-11$ orders of magnitude weaker than that of the strong D$_1$ or D$_2$ lines in cesium, for example.  Extracting the weak charge $Q_{\rm w}$ of the nucleus from the transition moment $\mathcal{E}_{\rm PNC}$ for the transition requires accurate theoretical models of the atomic wavefunctions.

The most precise value of $Q_{\rm w}$ in any atom to this point is derived from the APV measurements carried out by the Boulder group of Wieman~\cite{WoodBCMRTW97} in atomic cesium in 1997.  
One of the benefits of working in an alkali atom is its `simple' atomic structure, consisting of a single valence electron outside closed inner shells of electrons.  Models of the electronic wavefunctions of this heavy atom have become progressively more refined over the years~\cite{DzubaFS89, BlundellJS91, BlundellSJ92, Derevianko00, DzubaFG01, JohnsonBS01, KozlovPT01, DzubaFG02, FlambaumG05, PorsevBD09, PorsevBD10, DzubaBFR12, RobertsDF2013}.  To support and enhance these theoretical efforts, laboratory measurements of many atomic parameters, such as electric dipole (E1) matrix elements, have been carried out by several groups~\cite{BouchiatGP84, TannerLRSKBYK92, YoungHSPTWL94, RafacT98, RafacTLB99, BennettRW99, VasilyevSSB02, DereviankoP02a, AminiG03, BouloufaCD07, SellPEBSK11, ZhangMWWXJ13, antypas7p2013, Borvak14, PattersonSEGBSK15, GregoireHHTC15}. E1 matrix elements for transitions between low-lying levels are sensitive to the wavefunctions at moderate distances, comparable to the Bohr radius $a_0$, from the nucleus. The precision of many E1 matrix elements for transitions  between low-lying states of cesium is now $\sim 0.1$\%, and the experimental values are in very good agreement with theoretical values. (See Ref.~\cite{TohDTJE19} for a compilation of these results.)  

Since the weak-force interaction is a contact potential, calculations of $\mathcal{E}_{\rm PNC}$ also require precision in the wavefunctions near the nucleus.  
Theoretical efforts to calculate hyperfine coupling constants are therefore of great interest, since the hyperfine interaction is also sensitive to the electronic wavefunction at the nucleus. 
Calculations and measurements of the hyperfine coupling constants $A$, particularly of $J=1/2$ states, are therefore of critical importance to calculations of the weak Hamiltonian, and for gauging their precision.  

Recent theoretical work by Ginges \emph{et al.}~\cite{GingesVF17,GingesV18,GrunefeldRG2019} has focused on precision calculations of $A$ for low-lying and intermediate levels of atomic cesium.  Their relativistic Hartree-Foch many-body calculation includes effects of core polarization, correlation corrections, quantum electrodynamic (QED) radiative corrections (self-energy and vacuum polarization), and the Bohr-Weisskopf (BW) correction (an accounting of the nonuniform density of the magnetization of the nucleus).
In Ref.~\cite{GingesV18}, the authors proposed to use the results of precise measurements of the hyperfine splitting (hfs) of excited $ns \ ^2S_{1/2}$ states to greatly improve the ground $6s \ ^2S_{1/2}$ state and $7s \ ^2S_{1/2}$ state hyperfine intervals.  (Hereafter, the abbreviated notation $ns$ and $n p_J$ will be used in place of $ns \ ^2S_{1/2}$ and $np \ ^2P_{J}$, respectively.)  Their calculations showed that the correlation corrections decreased with increasing principal quantum number $n$, approaching a constant but non-zero value.  They proposed to use measurements of the hfs in high $ns$ states ($n>9$) to determine the BW and QED corrections in these states, which can then be scaled for application to the $6s$ and $7s$ states. This removes the large uncertainties due to the BW and QED corrections from the hfs calculations. We recently reported measurements~\cite{QuirkDTE22} of the hfs of the $12s$ and $13s$ states of cesium to be used for this analysis.  In Ref.~\cite{GrunefeldRG2019}, Grunefeld, Roberts, and Ginges examined trends in the corrections to the hyperfine coupling constants $A$, to make predictions of these constants for $ns$ and $np_{1/2}$ states of cesium, where $6 \le n \le 17$, which they believe to be accurate at the 0.1\% level. Recently, this group has found additional confirmation of the BW correction~\cite{SanamyanRG2} in historical data on muonic cesium.

Precise measurements of the hyperfine coupling constants of the $6p_{1/2}$ and $7p_{1/2}$ states of cesium have been reported previously.  The uncertainty in $A$ for these two states is 0.007\%~\cite{UdemRHH1999} and 0.04\%~\cite{FeiertagSzP1972,williamsHH2018}, respectively.  For the $8p_{1/2}$ state, however, the measurement uncertainty prior to the present measurement was $\sim0.2$\% \cite{TaiGH1973}. The goal of this study, therefore, was to determine the hyperfine coupling constant $A$ for the $8p_{1/2}$ state of cesium ($^{133}$Cs) with reduced uncertainty for direct comparison to the current theoretical results~\cite{GrunefeldRG2019}, and to facilitate improvements in these theoretical techniques. This measurement is in support of our ongoing investigation towards a high precision determination of the weak charge of atomic cesium~\cite{antypasm12013, choipnc2016, TohDTJE19}.  
In addition, we have examined the hyperfine structure of the $8p_{3/2}$ state, and report favorable comparison with prior experimental results, and determined the absolute energies of the $8p_{1/2}$ and $8p_{3/2}$ states with a precision of $\sim$150 kHz.

\section{\texorpdfstring{$8p \ ^2P_{1/2}$  measurements}{8p 2p1/2 and  measurements}\label{sec:8p12}}

\begin{figure}
    \centering
    \includegraphics[width=0.35\textwidth]{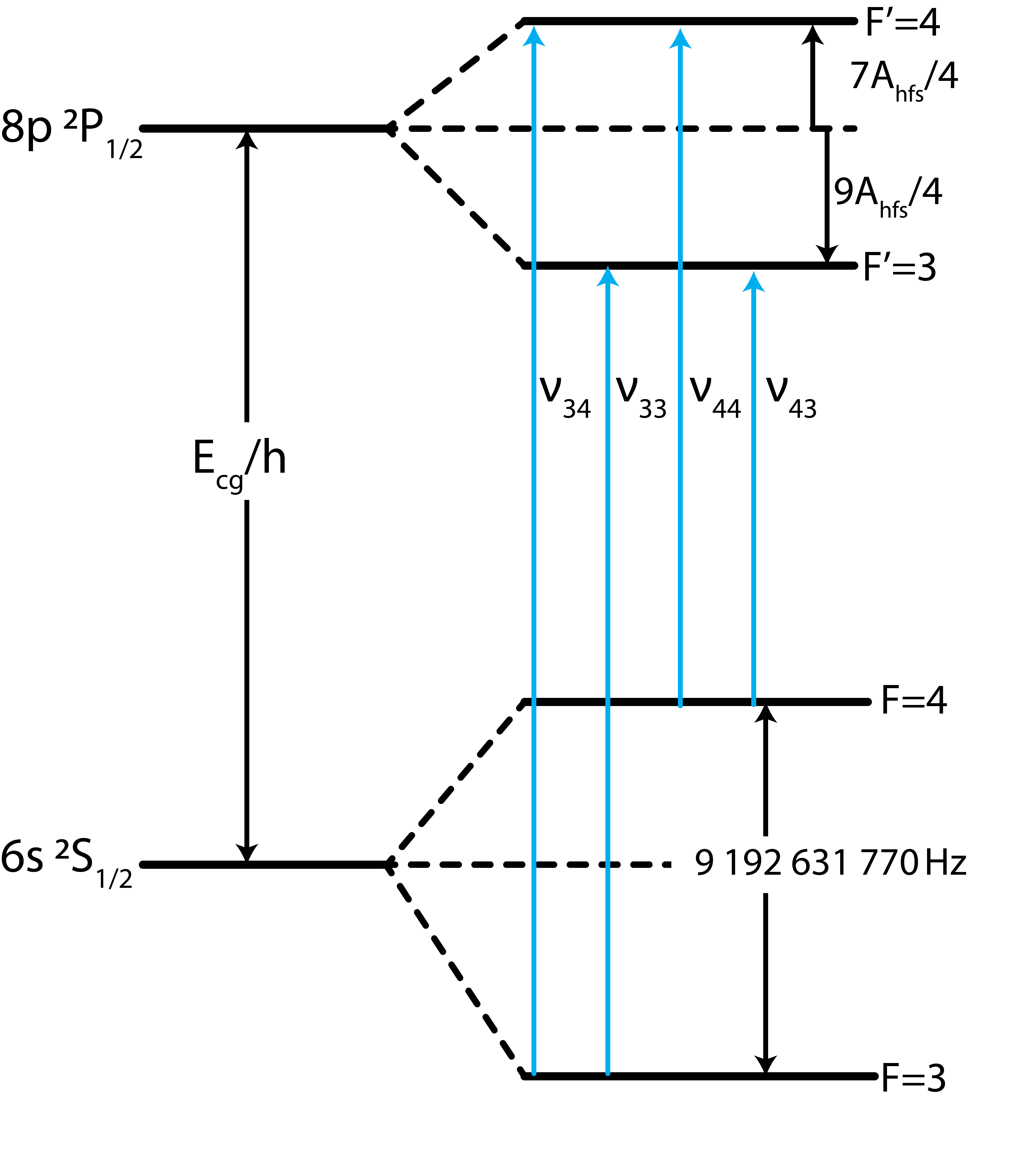}
    \caption{Energy level diagram showing the hyperfine components (not to scale) of the $6s$ and $8p_{1/2}$ states of cesium.  $\nu_{FF^{\prime}}$ indicates the frequency of the laser when resonant with the $6s,\; F \rightarrow 8p_{1/2},\; F^{\prime}$ electric dipole transition.  $E_{\rm cg}$ is the center of gravity energy of the $8p_{1/2}$ transition.}
    \label{fig:E_level_diagram}
\end{figure}

The energy levels of the ground $6s$ and excited $8p_{1/2}$ states of cesium are shown in Fig.~\ref{fig:E_level_diagram}. The hyperfine interaction splits both states ($6s$ and $8p_{1/2}$) into two hyperfine components, of energy (See Ref.~\cite{Corney1978atomic}),
\begin{eqnarray}\label{eq:Asplitting}
   E_{F=4} &=& E_{\rm cg} + \frac{7 h}{4} A  \\ 
   E_{F=3} &=& E_{\rm cg} - \frac{9 h}{4} A \nonumber
\end{eqnarray}
where $F$ is the total angular momentum (the vector sum of the nuclear $I=7/2$ and electronic $J=1/2$ angular momenta), $A$ is the relevant magnetic dipole hyperfine coupling constant. and $E_{cg}$ is the center-of-gravity energy of the state. The energy spacing of the ground state is defined to be $\Delta E_{\rm 6s}/h = 9.192 \: 631 \: 770$ GHz.

\subsection{Experimental Configuration and Procedure}\label{sec:Measurement1}
To measure the hyperfine splitting of the $8p_{1/2}$ state, we measure the absolute frequencies of the individual hyperfine components of the $6s \rightarrow 8p_{1/2}$ transitions.  To achieve this, we drive the electric-dipole allowed transition from the cesium ground state in an atomic beam using a cw narrow-band external cavity diode laser (ECDL), offset phase locked to a frequency comb laser (FCL) source.  
Precise frequency difference measurements can be made by referencing the driving laser's frequency to the FCL frequency.

 \begin{figure*}[t!]
\begin{centering}
\includegraphics[width=0.8\textwidth]{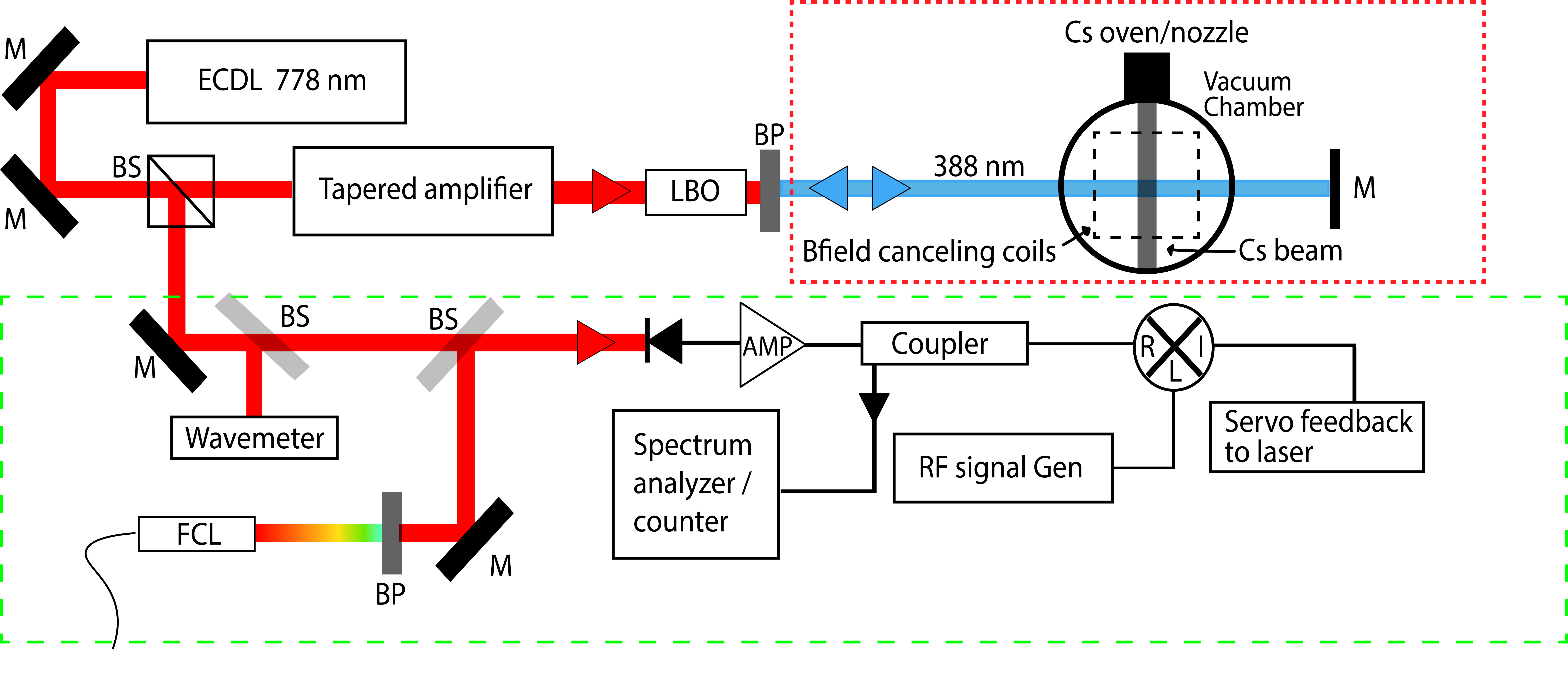}
   \caption{Experimental configuration for the $8p_{1/2}$ and $8p_{3/2}$ hyperfine spectroscopy. 3.5 W of 778 nm light is generated in a commercial ECDL and tapered amplifer unit and is focused into a lithium tri-borate crystal (LBO) in a single pass configuration to generate second harmonic light at 388-389 nm (170 $\mu$W) to excite the electric dipole transition. This doubled light is directed into a vacuum chamber, through an atomic beam and is retro-reflected to reduce Doppler shifts. A portion of the 778 nm light is beat against a frequency comb laser (FCL) and another portion is measured with a wavemeter. The combination of these two measurements yields precise absolute frequency measurements. We stabilize and narrow the light at 388 nm by offset phase locking the 778 nm laser to one of the comb teeth. This offset is varied to sweep the 388 nm light across the individual hyperfine levels. The following elements are label as; M - mirror, BP - band pass filter, BS - beam splitter, and LBO - lithium tri-borate crystal. The fine red dotted section includes the fluorescence detection and magnetic field canceling coils. The coarse green dotted section illustrates the frequency measurement and stabilization.}
 	  \label{fig:sweepsetup}
\end{centering} 	  
\end{figure*}

A schematic of the experimental configuration is illustrated in Fig.~\ref{fig:sweepsetup}. The 778 nm output of the commercial ECDL and tapered amplifier unit is frequency doubled in a lithium tri-borate (LBO) crystal in a single-pass geometry to produce $\sim$170 $\mu$W of light at 388.9 nm.  Fundamental light is separated from the second harmonic with a 40 nm wide band pass filter centered at 400 nm. This filter has an optical density of $>6.6$ at the fundamental frequency, and passes $96\%$ of the second harmonic. The 388.9 nm light is chopped (150 Hz) with a rotating chopper wheel and directed to a vacuum chamber where it crosses an atomic beam generated in an effusive oven. The light intersects the beam at close to a right angle and is retro-reflected to reduce Doppler shifts. Excited $8p$ atoms decay spontaneously via many pathways as they relax to the ground state.  The primary detection signal for our measurements is the 852 nm and 894 nm fluorescence of the $6p_{3/2} \rightarrow 6s$ and $6p_{1/2} \rightarrow 6s$ decay paths, respectively. 

The dimensions of the laser beam are $\sim$5 mm (width) $\times$ 2 mm (height).  
The aluminum vacuum chamber is cylindrical in shape, measuring 30 cm in diameter and 45 cm tall. It contains the oven, nozzle, photodetection system, and magnetic field biasing coils, and is pumped to a vacuum level of $5 \times 10^{-6}$ torr with a turbomolecular pump. The total magnetic field in the interaction region is reduced to below 10 mG. The atomic cesium beam is generated by a heated oven fitted with a nozzle composed of an array of stainless steel capillaries (0.58 mm I.D. x  $\sim$1 cm length). The beam then passes through a collimator, consisting of a stack of microscope coverslips (0.17 mm thick) spaced with microscope slides (1 mm thick), which restricts the beam divergence. 
A large area silicon photodiode and long pass ($>700$ nm) optical filter lie directly below the interaction region, which is defined by the intersection of the atomic beam and the laser beam. A curved reflector above the photodetector reflects upward fluorescence back down towards the photodetector. The long-pass filter effectively reduces scattered excitation light (388.9 nm), while efficiently passing longer-wavelength fluorescence (transmission $> 97$\%) at 852 nm ($6p_{3/2} \rightarrow 6s$) and 894 nm ($6p_{1/2} \rightarrow 6s$). The photo detection signal is amplified in a transimpedance amplifier and sent to a lock-in amplifier to be demodulated at the chopping frequency.

To stabilize the frequency of the ECDL, a portion of the light at 778 nm is beat against the output of the FCL. This source is a commercial femtosecond 1560 nm  fiber laser that is frequency doubled to 780 nm and spectrally broadened in a highly-nonlinear fiber.  The linewidth of each tooth of the comb laser source is less than one Hertz. The beat signal (at frequency $\nu_{\rm beat}$) is then fed into an analog optical-phase-lock loop. Here we amplify and mix down the beat signal with a stable signal generator to generate an error signal with which we lock the 778 nm source. By sweeping the frequency of the signal generator and counting the beatnote, we carefully control frequency scans across the $8p_{1/2}$ (and the $8p_{3/2}$ in Section \ref{sec:8p32}) spectra.

The absolute frequency of the second harmonic beam is given by 
\begin{equation}\label{eq:laserfreq}
\nu = 2(N \nu_{\rm rep} + \nu_{\rm offset} + \nu_{\rm beat}),
\end{equation}
where the factor of two accounts for exciting the transition with the second-harmonic light while beating the fundamental laser against the frequency comb, and $N$ represents the comb tooth number. $\nu_{\rm rep}$ and $\nu_{\rm offset}$ are the repetition rate and offset frequency of the FCL.  $N$ is determined using a wave meter with an accuracy of better than half of the repetition rate of the FCL ($\sim$250 MHz). The sign of the beat note is determined by observing the change in beat note while increasing the laser frequency.


We collect data in the following manner. After the temperature of the oven and nozzle have adequately stabilized to produce a consistent atomic beam density, the signal generator frequency is set to control the offset beatnote. The system pauses for a time $2\tau$, where $\tau = 100$ ms is the time constant of the lock-in amplifier. One hundred voltage samples are then collected using a 16-bit analog-to-digital converter (ADC) at a rate of one kHz, and are averaged and recorded. Ten sets of 100 voltage samples are collected. This protocol reduces correlation among the ten different data sets.  
The average and standard error of the mean of the ten voltage sets are then computed and recorded. The frequency of the signal generator is measured with a frequency counter and the beat signal itself is measured with a spectrum analyzer. Both of these frequencies are recorded. Then the signal generator is advanced to the next frequency. We collect a spectrum by stepping up and then back down through the optical transition and carefully search for drifts in the atomic beam density. A scan across the spectrum in both directions takes between four and six minutes, depending on the frequency width of the scan. We collect 15 to 20 spectra for each transition and fit the entire scan (both up and down-scans together). With these 15 to 20 individual fits, we determine the mean center frequency for each transition.

\subsection{Data Analysis}\label{sec:DataAnalysis1}
\begin{figure}
    \centering
    \includegraphics[width=0.4\textwidth]{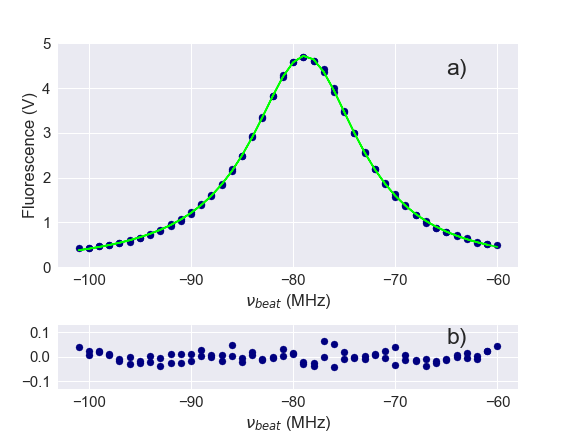}
    \caption{(a) A sample spectrum of a single hyperfine line, consisting of the fluorescence signal versus the beat frequency $\nu_{\rm beat}$. These data represent the $6s, F=3 \rightarrow 8p_{1/2}, F^{\prime} = 3$ line. Each point is a set of ten 100 ms (1 s total) measurements. The laser is stepped though 45 MHz and back in 1 MHz steps. The solid green line is the result of a least-squares fit of a Voigt function to the data.  (b) The residuals show the difference between the data points and the fitted function. }\label{fig:fluorescence8p1/2}
\end{figure}

We separately measure and record the absorption spectrum of each of the hyperfine components $6s, F \rightarrow 8p_{1/2}, F^{\prime}$, where $F=3,4$ ($F^{\prime}=3,4$) is the total angular momentum of the ground $6s$ (excited $8p_{1/2})$ state.
We show a single spectrum of the $6s, F=3 \rightarrow 8p_{1/2}, F^{\prime} = 3$ line, as a representative sample, in Fig.~\ref{fig:fluorescence8p1/2}(a).
The spectrum shows the fluorescence signal (lock-in amplifier output) versus the measured beat frequency between the fundamental (778 nm) laser and the nearest comb tooth of the frequency comb laser. 
The spectra are fit to a Voigt profile using a least-squares fitting algorithm. The fitting parameters include the amplitude, Gaussian and Lorentzian width, and center frequency of the peak, and a sloping baseline. The gently ($<1\%$ change) sloping baseline is produced by scattered light, and is present even in the absence of the atomic beam. 

The linewidth of this peak is primarily due to the divergence of the atomic beam.  Based on the geometry of our collimator and previous measurements with these instruments~\cite{GERGINOV200317}, we estimate an atomic beam divergence of $\sim$40 mrad. This is in excellent agreement with the measured linewidth of $\sim$26-27 MHz of the $6s \rightarrow 8p_{1/2}$ absorption peak. The natural lifetime for the $8p_{1/2}$ ($8p_{3/2}$) level is 376 ns (320 ns)~\cite{safronovadatabase_original}, resulting in a transition linewidth of 0.42 MHz (0.50 MHz). This homogeneous width could be increased slightly by power broadening, but still less than a MHz, as the laser intensity at the center of the beam is a factor of four below the saturation intensity for the $6s \rightarrow 8p_{1/2}$ transition for the highest power measurements.  (For the $6s \rightarrow 8p_{3/2}$ line, the greatest laser intensity used was comparable to the saturation intensity, so power broadening is somewhat larger, but still less than 1 MHz.)  Transit time broadening is expected to contribute less than 20 kHz, and collisional broadening less than 1 kHz.

The residuals (the difference between the data and the least-squares-fit result) are shown in Fig.~\ref{fig:fluorescence8p1/2}(b). This rms value of the residual is $\sim$0.5\% of the peak signal level, and is primarily due to thermal noise in the feedback resistor of the transimpedence amplifier.  

\begin{figure}
    \centering
    \includegraphics[width=0.45\textwidth]{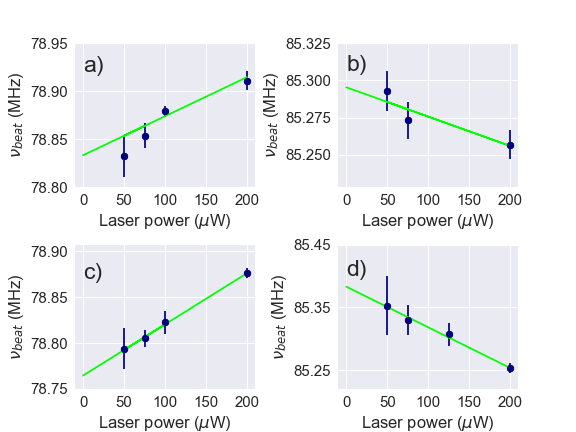}
    \caption{The power dependence of each of the $8p_{1/2}$ transitions. The dependence on the laser power ranged from 200-650 Hz/$\mu$W. (a) $6s,\: F=4\rightarrow 8p_{1/2},\: F'=4$, $\chi^2_{\rm red}=1.2$; (b) $6s,\: F=4\rightarrow 8p_{1/2},\: F'=3$, $\chi^2_{\rm red}=0.68$; (c) $6s,\: F=3\rightarrow 8p_{1/2},\: F'=4$, $\chi^2_{\rm red}=0.027$; (d) $6s,\: F=3\rightarrow 8p_{1/2},\: F'=3$, $\chi^2_{\rm red}=0.06$, where $\chi^2_{\rm red}$ is the reduced chi squared for the fit. In each plot, the blue dots are the data, and the green lines are the results of the least squares fit to the data. } \label{fig:8p_12_power}
\end{figure}

\subsection{Results}\label{sec:Results1}

To study the possible effect of Zeeman shifts, we intentionally apply a magnetic field of one Gauss. No magnetic field shifts or broadening were observed within the resolution of our measurement. Similarly, we studied the effect of a.c.\ Stark shifts by varying the laser power. A weak dependence (200-650 Hz/$\mu$W, varying among the lines) on the laser power was observed.  We corrected for this shift by fitting the measured linecenters vs.\ power to a linear function and extrapolating to zero laser power. This power dependence is illustrated in Fig.~\ref{fig:8p_12_power} for each of the hyperfine components of the $6s \rightarrow 8p_{1/2}$ transition. Using the zero-power extrapolated peak centers and Eq.~(\ref{eq:Asplitting}), we calculate the frequency difference between the hyperfine lines when driven from either of the $F=3$ or $4$ ground state hyperfine levels, and the hyperfine coupling constant ($A$) for the $8p_{1/2}$ level. These values of $A$ are 42.936(9) and 42.926(15) when driven from the $F=4$ and $F=3$ hyperfine level of the ground state, respectively. A weighted average of these two coupling constants is computed and presented in Table~\ref{table:8ponehalfhfsconstants}. This measurement of the hyperfine coupling constant ($A$) for the $8p_{1/2}$ state is in excellent agreement with previous measurements.  The 8 kHz uncertainty of our measurement is more than a factor of ten lower than the uncertainty of the previous measurement of Ref.~\cite{TaiGH1973}. This measurement agrees well with both theoretical values of Grunefeld \emph{et al.}, 2019~\cite{GrunefeldRG2019}. Results of two other theoretical calculations of $A$ for $8p_{1/2}$ are presented in Table~\ref{table:8ponehalfhfsconstants}; (Safronova \emph{et al.}, 1999~\cite{SafronovaJD99} and Tang \emph{et al.}, 2019~\cite{TangLS2019}).  Our measured value of $A$ differs from these results by $\sim$1-1.5\%.  

The sources of uncertainty for these measurements are recorded in Table ~\ref{table:sourcesoferror}. The uncertainty in the fit includes the statistical uncertainty in the repeated measurements of the center frequency, along with the uncertainty in extrapolating to zero laser power. The uncertainty in the frequency comb laser (FCL) is the uncertainty derived from the fractional uncertainty ($10^{-12}$) of the GPS conditioned time base used to stabilize the comb and the comb tooth number. The uncertainty in the shift due to the Zeeman effect is determined by the resolution of our measurement and the degree to which we cancel out magnetic fields. We observe no shifts in center frequency for any of the lines when applying a one Gauss field and we zero the magnetic field to better than 10 mGauss. With this, we estimate that the uncertainty is less than the resolution of the measurement times 1/100. The uncertainty due to beam misalignment is the residual Doppler error due to imperfect retro reflection. This uncertainty is only included in the absolute frequency determinations, discussed in Section ~\ref{sec:abs_freq}.

\begin{table}[t!] 
  \caption{Summary of results for the hyperfine coupling constant $A $, in MHz, of the $8p_{1/2}$ level. The numbers in parentheses following each value are the $1 \sigma$ standard error of the mean in the least significant digits.   }    \label{table:8ponehalfhfsconstants}
\begin{center}
 \begin{tabular}{cl} \hline \hline
 \rule{0in}{0.2in}$A$ & Source \\ \hline \multicolumn{2}{l}{Experiment\rule{0.2in}{0in}}   \\  
 \rule{0.1in}{0in}42.97 (10) \rule{0.1in}{0in} &   Tai \emph{et al.}, 1973~\cite{TaiGH1973}   \\ 
 42.92 (25)   &  Cataliotti \emph{et al.}, 1996~\cite{CataliottiFPI1996}  \\ 
 42.95 (25)  &   Liu \& Baird, 2000~\cite{LiuB2000} \\ 
42.933 (8)  &   This work \\ \hline
\multicolumn{2}{l}{Theory \rule{0.2in}{0in}}   \\  

 42.43  &   Safronova \emph{et al.}, 1999~\cite{SafronovaJD99} \\ 
 42.32  &  Tang \emph{et al.}, 2019~\cite{TangLS2019} \\ 
42.95 (9) &  fit method, Grunefeld \emph{et al.}, 2019~\cite{GrunefeldRG2019}\\
 42.93 (7)  &  ratio method, Grunefeld \emph{et al.}, 2019~\cite{GrunefeldRG2019} \\ \hline \hline
\end{tabular}
\end{center}
\end{table}

\begin{table}[b!]
    \caption{Sources of error and the uncertainty resulting from each, for the determinations of line centers for each of the spectra.   
    We add the errors in quadrature to obtain the total uncertainty. *Beam misalignment affects only the absolute frequency determinations.
    }
    \def\arraystretch{1.2}
    \begin{tabular}{|l|  c| }
        \hline

      \rule{0.1in}{0in}Source             &   \rule{0.1in}{0in}$\sigma_{\rm int}$(kHz)	\rule{0.1in}{0in}	\rule{0in}{0.15in}  \\
		\hline \hline
       \rule{0.1in}{0in} Fit, $\sigma_{\nu}$           	& 12-28   \\
		
       \rule{0.1in}{0in} FCL frequency, $\nu_{\rm FLC}$	& $<0.5$   \\
       \rule{0.1in}{0in} Zeeman			            	& $<0.2$ 	 \\
       \rule{0.1in}{0in} Beam misalignment*		            	& $145$ 	 \\

                                        \hline \hline
        \rule{0.1in}{0in}\rule{0in}{0.15in}Total Uncertainty, $\sigma_{\rm int}^{\rm total}$	\rule{0.1in}{0in}			& 12-28\\
        
    \hline
    \end{tabular}
    \label{table:sourcesoferror}
\end{table}

\section{\texorpdfstring{$8p \ ^2P_{3/2}$  measurements}{8p 2p3/2 and  measurements}\label{sec:8p32}}

\begin{figure}
    \centering
    \includegraphics[width=0.35\textwidth]{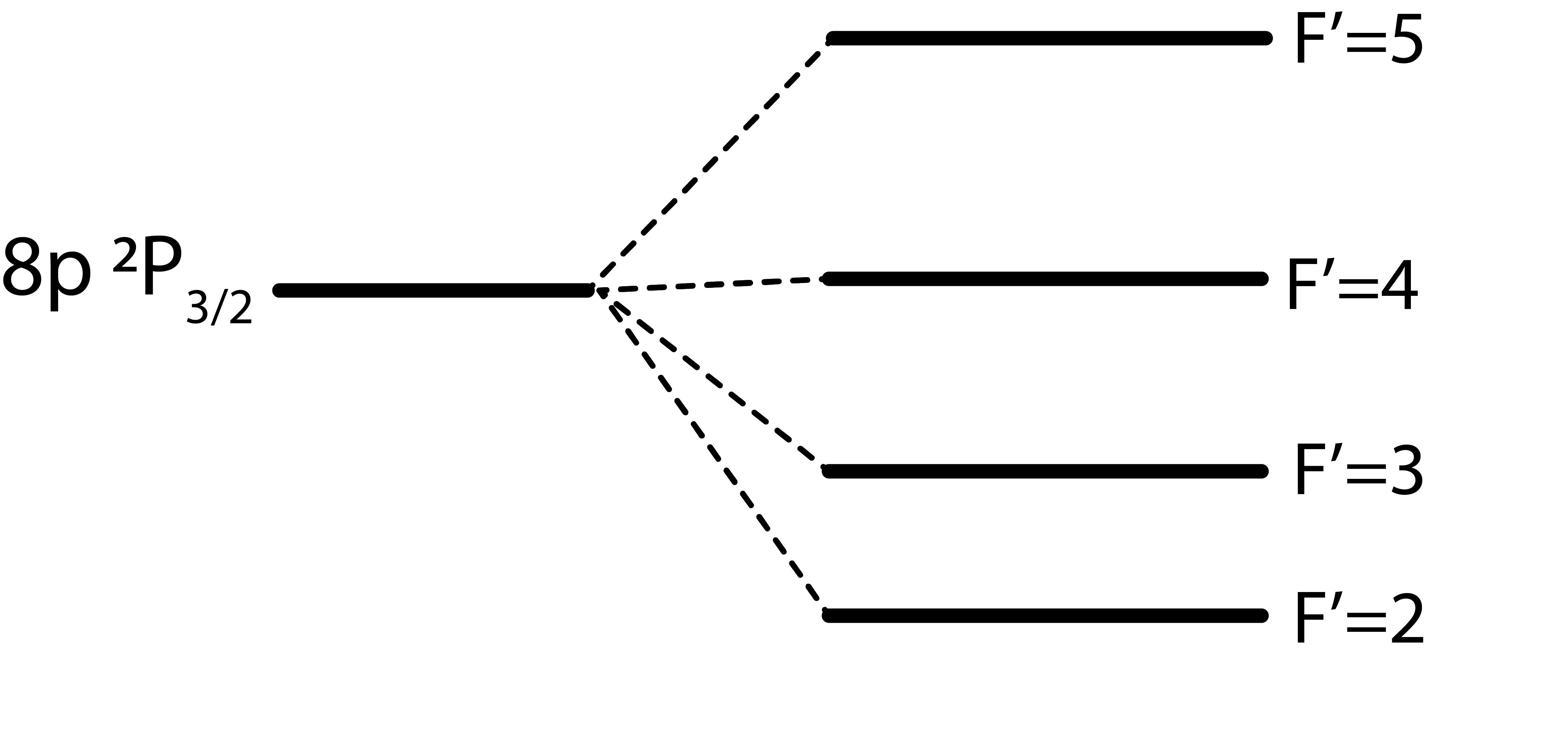}
    \caption{Energy level diagram showing the hyperfine components (not to scale) of the $8p_{3/2}$ states of cesium. }
    \label{fig:E_level_diagram_8p32}
\end{figure}

The energy levels of the $8p_{3/2}$ states are shown in Fig.~\ref{fig:E_level_diagram_8p32}.
The energies of the hyperfine components of the $8p \: ^2P_{3/2}$ state are shifted from the center-of-gravity energy $E_{\rm cg}$ of the 8p state by the magnetic dipole ($A$), electric quadrupole ($B$), and magnetic octupole ($C$) interactions.  The coupling constants for these interactions are represented by $A$, $B$, and $C$, as indicated. The energy shifts due to each of these terms are
\begin{eqnarray}\label{eq:levelenergy3halves}
   E_{F^{\prime}=5} &=& E_{\rm cg} + \frac{h}{4} \left(21 A + B + 4C \right)  \nonumber \\ 
   E_{F^{\prime}=4} &=& E_{\rm cg} + \frac{h}{28} \left(7A -13B -132C \right)  \\
   E_{F^{\prime}=3} &=& E_{\rm cg} + \frac{h}{28} \left(-105A -5B +220C \right)\nonumber \\
   E_{F^{\prime}=2} &=& E_{\rm cg} + \frac{h}{28} \left(-189A +15B -132C \right)\nonumber
\end{eqnarray}
as given in Refs.~\cite{Corney1978atomic, GerginovDT2003}.  The spacing between lines then yield the hyperfine coupling constants $A$, $B$, and $C$, using 
\begin{eqnarray}\label{eq:ABC_from_DelNu}
   A = \frac{11}{120}\Delta \nu_{54} + \frac{2}{21}\Delta \nu_{43} + \frac{3}{56}\Delta \nu_{32} \nonumber \\
   B = \frac{77}{120}\Delta \nu_{54} - \frac{1}{3}\Delta \nu_{43} - \frac{5}{8}\Delta \nu_{32}  \\
   C = \frac{7}{480}\Delta \nu_{54} - \frac{1}{24}\Delta \nu_{43} + \frac{1}{32}\Delta \nu_{32}, \nonumber 
\end{eqnarray}
where $\Delta \nu_{ij} = (E_{F^{\prime}=i} - E_{F^{\prime}=j} )/h$ is the frequency spacing between the hyperfine peaks.

\subsection{Experimental Configuration and Procedure}\label{sec:Measurement3}
\begin{figure}
    \centering
    \includegraphics[width=0.5\textwidth]{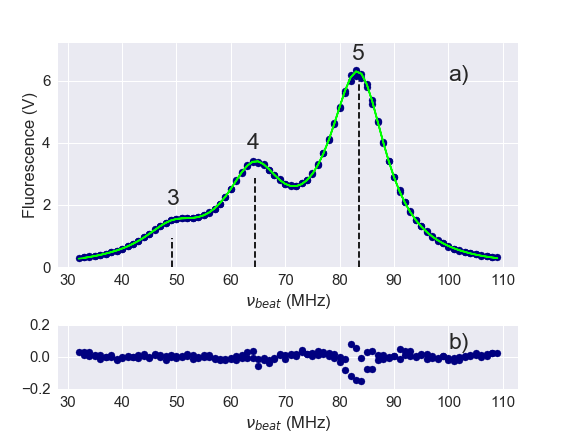}
    \caption{(a) Spectra of the $6s \: F=4 \rightarrow 8p_{3/2}, \: F'=n$ transitions where $n \in \{3,4,5\}$. The blue points are measured fluorescence and the green trace is the result of a least-squares fit to the measured fluorescence. The dotted vertical lines indicate the fitted peak centers and the calculated relative line strengths of each transition. (b) Residuals of the least squares fit.} \label{fig:8p_32_f4}
\end{figure}

\begin{figure}
    \centering
    \includegraphics[width=0.5\textwidth]{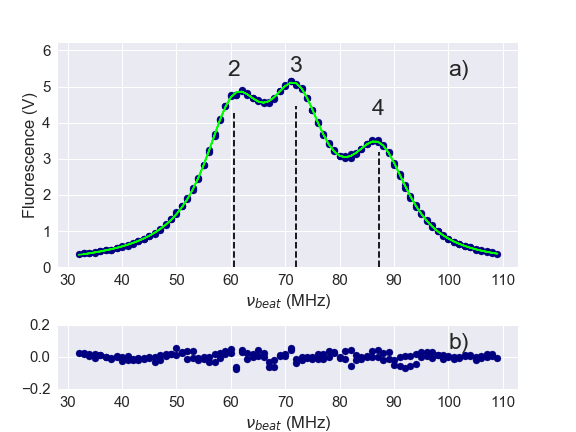}
    \caption{(a) Spectra of the $6s \:F=3 \rightarrow 8p_{3/2},\: F'=n$ transitions where $n \in \{2,3,4\}$ The blue points are measured fluorescence and the green trace is a least-squares fit to the measured fluorescence. The dotted vertical lines indicate the fitted peak centers and the calculated relative line strengths of each transition. (b) Residuals of the least squares fit.} \label{fig:8p_32_f3}
\end{figure}

The experimental procedure for measuring the hyperfine splitting on the $8p_{3/2}$ line is similar to that for the $8p_{1/2}$, but varies in a few ways. The wavelength of the transition is similar, 387.7 nm, so the same laser source and LBO crystal are used.  The hyperfine splitting on the $8p_{3/2}$ line is less than that of the $8p_{1/2}$ line, allowing us to scan across all of the allowable transitions in a single sweep. The required scan length is larger by 1.5 to 2 times to include each peak in a single scan. Unfortunately, this smaller spectral spacing also means that the individual lines are not completely resolvable (see Figs.~\ref{fig:8p_32_f4} and ~\ref{fig:8p_32_f3}). The final difference between the two procedures is that the transition strength of the $8p_{3/2}$ line is larger than that of the $8p_{1/2}$ line by a factor of 5-10. This allows us to lower the power in the second harmonic beam driving the $6s\rightarrow 8p_{3/2}$ transition without sacrificing the signal-to-noise ratio.

\subsection{Data Analysis}\label{sec:DataAnalysis3}
Examples of $8p_{3/2}$ hyperfine spectra driven from the $F=4$ and $F=3$ hyperfine ground states are illustrated in Figs. \ref{fig:8p_32_f4} and \ref{fig:8p_32_f3}.  We fit the $8p_{3/2}$ spectra with the sum of three Voigt profiles. The fitting parameters include the center frequency for each peak, a single Gaussian and Lorentzian width (all peaks were constrained to the same widths), amplitudes for each peak, and a sloping baseline. The peaks were also allowed to have a slight asymmetry to account for imperfect alignment through the atomic beam. The relative peak heights of the various hyperfine components in spectra such as Figs. 6 or 7 are within $10\%$ of calculated values based on Clebsch-Gordon coefficients at the lowest optical intensities.

\subsection{Results}\label{sec:Results3}
We investigated a.c.\ Stark and Zeeman shifts on the $8p_{3/2}$ lines. Within the resolution of the measurement, no observable shifts were observed when applying a 1 Gauss magnetic field. When varying the power from $50 \: \mu$W to $170 \: \mu$W, a slight shift in hyperfine peak spacing was observed.  The power dependence of each of the hyperfine lines varies in the range of 200-700 Hz/$\mu$W. We fit these individual splittings versus power and extrapolate to zero laser intensity. These values are reported in Table ~\ref{table:8p32_levelsplittings}.  Two values of $\Delta \nu_{43}$ are reported here, one for excitation out of the $F=3$ component of the ground state, the other for excitation out of the $F=4$ ground state.  These values differ by somewhat more than their combined uncertainties.  We use the weighted average of these two values, with the uncertainty expanded by $\sqrt{\chi^2_{\rm red}} = 2$~\cite{bevington2003}.

We take the fitted zero power hyperfine splittings from Table ~\ref{table:8p32_levelsplittings} and calculate hyperfine coupling constants $A$, $B$, and $C$ using Eq.\ (\ref{eq:ABC_from_DelNu}). A summary of these results and those of previous measurements are reported in Table~\ref{table:8p3halveshfsconstants}. Our value for $A$ is in fair agreement with measurements of Refs.~\cite{Bucka1962} and \cite{RydbergS1972}; less so with those of Refs.~\cite{FaistGK1964,AbeleBH1975,BayramAPGBHM2014}. The uncertainty of our measurement is comparable to, but slightly larger than that of Refs.\ \cite{Bucka1962} and \cite{RydbergS1972}. Prior values of the electric quadrupole coupling constant $B$ from previous measurements have not been consistent, with uncertainties comparable to the values themselves. Our result of $B=-0.005 \: (40)$ MHz is less than its uncertainty.  Remarkably, our measurements lead to a value of the magnetic octupole constant $C$ for this level.  There have been no previous reports of this constant for the $8p_{3/2}$ state.

\begin{table}[t!]
\begin{center}
 \def\arraystretch{1.3}
 \begin{tabular}{|c|c|c|c|} \hline
\rule{0in}{0.1in}   & \multicolumn{3}{c|}{\rule{0in}{0.15in}Fitted values}  \\ \cline{2-4} 
Initial State & $\Delta \nu_{32}$ (MHz) &  $\Delta \nu_{43}$ (MHz)   &  $\Delta \nu_{54}$ (MHz) \\  \hline
\rule{0.00in}{0.0in}$F=3$&23.026(38)&  30.120(74) &- \\
 
\rule{0.00in}{0in}$F=4$& -&   30.298(50) & 38.130(30) \\
\rule{0.0in}{0in}Combined    & 23.026(38)&  30.242(82)  &38.130(30)
\\ \hline \hline 
\end{tabular}
\end{center} 
  \caption{A summary of the results of the $8p_{3/2}$ hyperfine splittings. The fitted values are the fitted zero intensity splittings.  The numbers in parentheses following each value are the $1 \sigma$ standard error of the mean in the least significant digits. }  \label{table:8p32_levelsplittings}
\end{table}

\begin{table}[t!] 
  \caption{Summary of results for the hyperfine coupling constants $A$, $B$, and $C$, in MHz, of the $8p_{3/2}$ level. The numbers in parentheses following each value are the $1 \sigma$ standard error of the mean in the least significant digits.   }    \label{table:8p3halveshfsconstants}
\begin{center}
 \begin{tabular}{llll} \hline \hline
 \rule{0in}{0.2in}$A$ & $B$ & $C$ & Source \\ \hline
 \multicolumn{4}{l}{\emph{Experiment}\rule{0in}{0.2in}}   \\  
\rule{0in}{0.15in}7.626 (5) & -0.049 (42) &   \rule{0in}{0.15in} &   Bucka \emph{et al.}, 1962 \cite{Bucka1962}  \\ 
\rule{0in}{0.15in}7.58 (1) & -0.14 (5) &  & Faist \emph{et al.}, 1964 ~\cite{FaistGK1964} \\ 
\rule{0in}{0.15in}7.626 (5)  &  -0.090 (24) &  \rule{0in}{0.15in} &  Rydberg \emph{et al.}, 1972 \cite{RydbergS1972} \\ 
\rule{0in}{0.15in}7.644 (25) &  &  & Abele \emph{et al.}, 1975~\cite{AbeleBH1975} \\
\rule{0in}{0.15in}7.42 (6) &  \rule{0in}{0.15in} 0.14 (29) &  & Bayram \emph{et al.}, 2014~\cite{BayramAPGBHM2014} \\
\rule{0in}{0.15in}7.609 (8) & -0.005 (40) & 0.016 (4) & This work, 2022  \\

\multicolumn{4}{l}{\emph{Theory} \rule{0in}{0.2in}}\\  

\rule{0in}{0.15in}7.58 (5) & -0.046 (35) &   &  Barbey \emph{et al.}, 1962 \cite{osti_4737265} \\ \rule{0in}{0.15in}7.27 &  & &  Safronova \emph{et al.}, 1999~\cite{SafronovaJD99} \\  \rule{0in}{0.15in}7.44 &   &  & Tang \emph{et al.}, 2019~\cite{TangLS2019} \\
\\ \hline \hline
\end{tabular}
\end{center}
\end{table}

\section{\texorpdfstring{Absolute frequency measurements}{Absolute frequency measurements}\label{sec:abs_freq}}
Along with high precision determinations of the hyperfine coupling constants for the $8p_{1/2}$ and $8p_{3/2}$ states, our measurements also provide high precision absolute frequency determinations for these states. We use the fitted peak centers from each of the hyperfine lines in Eqs.~(\ref{eq:Asplitting}) and (\ref{eq:levelenergy3halves}) to calculate the center of gravity frequency for both the $6s\rightarrow 8p_{1/2}$ and $6s\rightarrow 8p_{3/2}$ transitions, respectively.  Due to possible beam misalignment, we estimate the uncertainty in the linecenter frequency to be 145 kHz, which is added in quadrature with the respective line center uncertainty due to the fitting procedure, the frequency comb uncertainty, and the uncertainty in the Zeeman shift. We report these absolute frequency measurements in Table~\ref{table:absolutefrequency}. Our values for the line center frequencies agree well with the measurements by Kleiman \cite{Kleiman:62} for both lines.  The uncertainty of our measurement, however, is significantly smaller. The value for the $6s\rightarrow 8p_{1/2}$ linecenter frequency by Liu and Baird \cite{LiuB2000} is in poor agreement with our measurement and that of Kleiman.

\begin{table}[t!] 
  \caption{Summary of results for the absolute frequency measurements of the $8p_{1/2}$ and $8p_{3/2}$ levels. The numbers in parentheses following each value are the $1 \sigma$ standard error of the mean in the least significant digits.   }    \label{table:absolutefrequency}
\begin{center}
 \begin{tabular}{cl} \hline \hline
 \rule{0in}{0.15in} \rule{0.2in}{0in} Line center \rule{0.2in}{0in} & \\  \rule{0.2in}{0in} frequency (MHz)\rule{0.2in}{0in} & Source \\ \hline \multicolumn{2}{l}{$8p_{1/2}$ }   \\  
 
 $770 \: 731 \: 690 \: (150)$   &  Kleiman, 1962~\cite{Kleiman:62}  \\ 
 $770 \: 731 \: 498.0 \: (10)$  &   Liu \& Baird, 2000~\cite{LiuB2000} \\ 
$770 \: 731 \: 653.30  \: (15)$  &   This work \\ \hline
\multicolumn{2}{l}{$8p_{3/2}$}   \\  

$773 \: 210 \: 080 \: (150)$ &  Kleiman, 2019~\cite{GrunefeldRG2019} \\ 
 $773 \: 210 \: 182.30 \: (15)$  &   This work  \\ \hline \hline
\end{tabular}
\end{center}
\end{table}

\section{Conclusion}\label{sec:Conclusions}

In this work, we have reported a new, high precision measurement of the hyperfine coupling constant $A = 42.933 \: (8)$ for the $8p_{1/2}$ state in atomic cesium-133. This value is in support of theoretical efforts towards high precision values of electronic wavefunctions, which are a critical component in characterizing the parity nonconserving weak interaction. We also report values for the hyperfine coupling constants $A$, $B$, and $C$ for the $8p_{3/2}$ state as well as absolute frequency measurements of both the $8p_{1/2}$ and $8p_{3/2}$ states.

This material is based upon work supported by the National Science Foundation under grant number PHY-1912519 and the REU grant number PHY-1852501. We acknowledge useful conversations with J.~Ginges, and frequency comb laser advice from D.~Leaird and N.~O'Malley.

\bibliography{biblio}

\end{document}